\documentclass[12pt]{article}

\newcommand{\uDelta}{\Delta}
\newcommand{\uPi}{\mathrm{P}}
\newcommand{\uLambda}{{\Lambda}}
\newcommand{\A}{\mathbf{A}}
\newcommand{\B}{\mathbf{B}}
\newcommand{\x}{\mathbf{x}}
\newcommand{\bk}{\mathbf{k}}
\newcommand{\R}{\mathbf{R}}
\newcommand{\s}{\boldsymbol{\sigma}}
\newcommand{\p}{\mathbf{p}}
\newcommand{\da}{\boldsymbol{\alpha}}
\newcommand{\D}{\mathrm{d}}
\usepackage{amstex}   % useful for coding complex math

\begin{document}
\title{The Stability of Matter and Quantum Electrodynamics}
%
%
%\toctitle{The Stability of Matter and Quantum Electrodynamics}
% allows explicit linebreak for the table of content
%
%
%\titlerunning{Stability of Matter and QED}
% allows abbreviation of title, if the full title is too long
% to fit in the running head
%
\author{Elliott H. Lieb}

\date{September 2, 2002}
%
%\authorrunning{Elliott H. Lieb}
% if there are more than two authors,
% please abbreviate author list for running head
%
%
\maketitle              % typesets the title of the contribution

\renewcommand{\thefootnote}{}

\renewcommand{\thefootnote}{}
\footnotetext{For publication in the proceedings of
the Werner Heisenberg Centennial, Munich, December, 2001.}
\renewcommand{\thefootnote}{}
\footnotetext{
\copyright\, 2002 by the author. This article may be reproduced, in its
entirety, for non-commercial purposes.}
\renewcommand{\thefootnote}{}
\footnotetext{Work partially
supported by U.S. National Science Foundation
grant PHY 0139984.}
%
%%%%%%%%%%%%%%%%%%%%%%%%%%%%%%%%%%%%%%%%%%%%%%%%%%%%%%%%%%%%%

\section{Foreword}

Heisenberg was undoubtedly one of the most important physicists of the
20th century, especially concerning the creation of quantum mechanics.
It was, therefore, a great honor and privilege for me to be asked to
speak at this symposium since quantum mechanics is central to my own
interests and forms the basis of my talk, which is about the quantum
theory of matter in the large and its interaction with the quantized
radiation field discovered earlier by Planck.

My enthusiastic participation in the scientific part of this symposium
was tempered by other concerns, however. Heisenberg has become, by
virtue of his importance in the German and world scientific community,
an example of the fact that a brilliant scientific and highly cultured
mind could coexist with a certain insensitivity to political matters
and the way they affected life for his fellow citizens and
others. Many opinions have been expressed about his participation in
the struggle of the Third Reich for domination, some forgiving and
some not, and I cannot judge these since I never met the man.  But
everyone is agreed about the fact that Heisenberg could view with
equanimity, if not some enthusiasm, the possibility of a German
victory, which clearly would have meant the end of 
civilization as we know it and enjoy it. By the start of the war this
fact was crystal clear, or should have been clear if humanistic
culture has more than a superficial meaning.  To me it continues to be
a mystery that the same person could see the heights of human culture
and simultaneously glimpse into the depths of depravity and not see
that the latter would destroy the former were it not itself destroyed.

\section{Introduction}

The quantum mechanical revolution brought with it many  successes but
also a few problems that have yet to be resolved. We begin with a sketch
of the topics that will concern us here.

\subsection{Triumph  of Quantum Mechanics}\label{triumph}

One of the basic problems of classical physics (after the discovery of
the point electron by Thompson and of the (essentially) point nucleus
by Rutherford) was the stability of atoms. Why do the electrons in an
atom not fall into the nucleus? Quantum mechanics explained this fact.
It starts with the classical Hamiltonian of the system (nonrelativistic
kinetic energy for the electrons plus Coulomb's law of electrostatic
energy among the charged particles). By virtue of the non-commutativity
of the kinetic and potential energies in quantum mechanics the stability
of an atom -- in the sense of a finite lower bound to the energy -- was a
consequence of the fact that any attempt to make the electrostatic energy
very negative would require the localization of an electron close to the
nucleus and this, in turn, would result in an even greater, positive,
kinetic energy.

Thus, the basic stability problem for an atom was solved by an inequality
that says that $\langle 1/|x| \rangle$ can be made large only at the
expense of making $\langle p^2 \rangle$ even larger. In elementary
presentations of the subject it is often said that the mathematical
inequality that ensures this fact is the famous uncertainty principle of
Heisenberg (proved by Weyl), which states that $\langle p^2 \rangle 
\langle x^2 \rangle
\geq (9/8)\hbar^2 $ with $\hbar =h/2\pi$ and $h=$Planck's constant.

While this principle is mathematically rigororous it is actually
insufficient for the purpose, as explained, e.g., in
\cite{lieb1,lieb2}, and thus gives only a heuristic explanation of the
power of quantum mechanics to prevent collapse.  A more powerful
inequality, such as Sobolev's inequality (\ref{sobolev}), is needed
(see, e.g., \cite{anal}). The utility of
the latter is made possible by Schr\"odinger's representation of
quantum mechanics (which earlier was a somewhat abstract theory of
operators on a Hilbert space) as a theory of differential operators on
the space of square integrable functions on $\mathbb{R}^3$. The
importance of Schr\"odinger's representation is sometimes
underestimated by formalists, but it is of crucial importance because
it permits the use of functional analytic methods, especially
inequalities such as Sobolev's, which are not easily visible on the
Hilbert space level. These methods are essential for the developments
reported here.

To summarize, the understanding of the stability of atoms and ordinary
matter requires a formulation of quantum mechanics with two ingredients:
\noindent \begin{itemize} \item A Hamiltonian formulation in order to
have a clear notion of a lowest possible energy. Lagrangian formulations,
while popular, do not always lend themselves to the identification of
that quintessential quantum mechanical notion of a ground state energy.
\item A formulation in terms of concrete function spaces instead of
abstract Hilbert spaces so that the power of mathematical analysis can
be fully exploited.  \end{itemize}

\subsection{Some Basic Definitions}

As usual, we shall denote the lowest energy (eigenvalue) of a quantum
mechanical system by $E_0$. (More generally, $E_0$ denotes the infimum
of the spectrum of the Hamiltonian $H$ in case this infimum is not an
eigenvalue of $H$ or is $-\infty$.) Our intention is to investigate
arbitrarily large systems, not just atoms. In general we suppose that
the system is composed of $N$ electrons and $K$ nuclei of various kinds.
Of course we could include other kinds of particles but $N$ and $K$
will suffice here. $N=1$ for a hydrogen atom and $N=10^{23}$ for a mole of
hydrogen.  We shall use the following terminology for two notions of stability:
\begin{eqnarray} \label{firststab}
&E_0&
> -\infty  \quad\qquad\qquad\qquad \mathrm{Stability\ of\  the\  first\
kind,}\\ 
&E_0& > C(N+K)  \qquad\qquad \mathrm{Stability\
of\  the\  second\  kind} \label{secondstab}
\end{eqnarray} 
for some constant $C\leq 0$ that is independent of $N$ and $K$, but
which may depend on the physical parameters of the system (such as the
electron charge and mass). Usually, $C<0$, which means that there is a
positive binding energy per particle.

Stability of the second kind is absolutely essential if quantum mechanics
is going to reproduce some of the basic features of the ordinary material
world: The energy of ordinary matter is extensive, the thermodynamic
limit exists and the laws of thermodynamics hold. Bringing two stones
together might produce a spark, but not an explosion with a release of
energy comparable to the energy in each stone.  Stability of the second
kind does not guarantee the existence of the thermodynamic limit for
the free energy, but it is an essential ingredient \cite{lieblebowitz} 
\cite[Sect. V]{lieb1}.

It turns out that stability of the second kind cannot be taken for
granted, as Dyson discovered \cite{dyson1}. If Coulomb forces are
involved, then {\it the Pauli exclusion principle is essential.}
Charged bosons are {\it not stable} because for them $E_0\sim -N^{7/5}$
(nonrelativistically) and $E_0 = -\infty$ for large, but finite $N$ 
(relativistically, see Sect.
\ref{relmanybody}).

\subsection{The Electromagnetic Field}

A second big problem handed down from classical physics was
the `electromagnetic mass' of the electron. This poor creature has
to drag around an infinite amount of electromagnetic energy that
Maxwell burdened it with. Moreover, the electromagnetic field itself
is quantized -- indeed, that fact alone started the whole revolution.

While quantum mechanics accounted for stability with Coulomb forces
and Schr\"odinger led us to think seriously about the `wave function
of the universe', physicists shied away from talking about the wave
function of the particles in the universe {\it and} the
electromagnetic field in the universe. It is noteworthy that
physicists are happy to discuss the quantum mechanical many-body
problem with external electromagnetic fields non-perturbatively, but
this is rarely done with the quantized field. The quantized field
cannot be avoided because it is needed for a correct description of
atomic radiation, the laser, etc. However, the interaction of matter
with the quantized field is almost always treated perturbatively or
else in the context of highly simplified models (e.g., with two-level
atoms for lasers).

The quantized electromagnetic field greatly complicates the stability
of matter question. It requires, ultimately, mass and charge
renormalizations.  At present such a complete theory does not exist,
but a theory {\it must} exist because matter exists and because we
have strong experimental evidence about the manner in which the
electromagnetic field interacts with matter, i.e., we know the
essential features of a low energy Hamiltonian.  In short, nature
tells us that it must be possible to formulate a self-consistent
quantum electrodynamics (QED) {\it non-perturbatively,} (perhaps with
an ultraviolet cutoff of the field at a few MeV).  It should not be
necessary to have recourse to quantum chromodynamics (QCD) or some
other high energy theory to explain ordinary matter.  

Physics and other natural sciences are successful because physical
phenomena associated with each range of energy and other parameters
are explainable to a good, if not perfect, accuracy by an appropriate
self-consistent theory. This is true whether it be hydrodynamics,
celestial dynamics, statistical mechanics, etc.  If low energy physics
(atomic and condensed matter physics) is not explainable by a
self-consistent, non-perturbative theory on its own level one can
speak of an epistemological crisis.

Some readers might say that QED is in good shape. After all, it
accurately predicts the outcome of some very high precision experiments
(Lamb shift, $g$-factor of the electron). But the theory does not
really work well when faced with the problem, which is explored here,
of understanding the many-body ($N\approx 10^{23}$) problem and the
stable low energy world in which we spend our everyday lives.

\subsection{Relativistic Mechanics}

When the classical kinetic energy $p^2/2m$ is replaced by its relativistic
version $\sqrt{p^2c^2 +m^2c^4} $ the stability question becomes
much more complicated, as will be seen later. It turns out that even
stability of the first kind is not easy to obtain and it depends on the
values of the physical constants, notably the fine structure constant
\begin{equation} \label{alpha}
\alpha= e^2/\hbar c =1/137.04 \ ,
\end{equation} 
where $-e$ is the
electric charge of the electron.

For ordinary matter relativistic effects are not dominant but they are
noticeable. In large atoms these effects severely change the innermost
electrons and this has a noticeable effect on the overall electron
density profile.  Therefore, some version of relativistic mechanics is
needed, which means, presumably, that we must know how to replace
$p^2/2m$ by the Dirac operator.

The combination of relativistic mechanics plus the electromagnetic field
(in addition to the Coulomb interaction) makes the stability problem
difficult and uncertain.  Major aspects of this problem have been worked
out in the last few years (about 35) and that is the subject of this lecture.

\section{Nonrelativistic Matter without the Magnetic Field}\label{nomagnet}

We work in the `Coulomb' gauge for the electromagnetic field. Despite
the assertion that quantum mechanics and quantum field theory are
gauge invariant, it seems to be essential to use this gauge, even
though its relativistic covariance is not as transparent as that of the
Lorentz gauge. The reason is the following.

In the Coulomb gauge the electrostatic part of the interaction of matter with 
the electromagnetic field is put in `by hand', so to speak. That is, it is 
represented by an ordinary potential $\alpha V_c$, of the form
(in energy units $mc^2$ and length units the Compton wavelength $\hbar/mc$)
\begin{eqnarray}
V_c = - \sum_{i=1}^N \sum_{k=1}^K {Z_k\over |\x_i - \R_k|}
+\sum_{1\leq i < j \leq N}{1\over |\x_i-\x_j|} 
+\sum_{1\leq k < l \leq K}{Z_kZ_l\over|\R_k-\R_l|} \  .
\end{eqnarray}
The first sum is the interaction of the electrons (with dynamical
coordinates $\x_i$) and fixed nuclei located at $\R_k$ of positive
charge $Z_k$ times the (negative) electron charge $e$. The second is
the electron-electron repulsion and the third is the nucleus-nucleus
repulsion. The nuclei are fixed because they are so massive relative
to the electron that their motion is irrelevant.  It could be
included, however, but it would change nothing essential.  Likewise
there is no nuclear structure factor because if it were essential for
stability then the size of atoms would be $10^{-13}$ cm instead of
$10^{-8}$ cm, contrary to what is observed. 

Although the nuclei are fixed the constant $C$ in the stability of
matter (\ref{secondstab}) is required to be independent of the $\R_k$'s.
Likewise (\ref{firststab}) requires that $E_0$ have a finite lower bound
that is independent of the $\R_k$'s.

For simplicity of exposition we shall assume here that all the 
$Z_k$ are identical, i.e., $Z_k=Z$.

The magnetic field, which will be introduced later, is described by a 
vector potential $\A(x)$ which is a dynamical variable in the Coulomb gauge.
The magnetic field is $\B=\mathrm{curl}\A$.

There is a basic physical distinction between electric and magnetic forces
which does not seem to be well known, but which motivates this choice of gauge.
In electrostatics like charges repel while in magnetostatics like currents
attract. A consequence of these facts is that the correct magnetostatic
interaction energy can be obtained by minimizing the energy functional
$\int B^2 + \int \mathbf{j}\cdot \A$ with respect to the vector field $\A$. 
The electrostatic energy, on the other hand, {\it cannot} be obtained by a 
minimization principle with respect to the field (e.g., 
minimizing $\int | \boldsymbol{\nabla} \phi|^2 + 
\int \phi \varrho$ with respect to $\phi$). 

The Coulomb gauge, which puts in the electrostatics correctly, by
hand, and allows us to minimize the total energy with respect to the
$\A$ field, is the gauge that gives us the correct physics and is
consistent with the ``quintessential quantum mechanical notion of a
ground state energy'' mentioned in Sect. \ref{triumph}. In any other
gauge one would have to look for a critical point of a Hamiltonian
rather than a true global minimum.

The type of Hamiltonian that we wish to consider in this
section is
\begin{equation}
H_N= T_N+ \alpha V_c\ . 
\end{equation}
Here, $T$ is the kinetic energy of the $N$ electrons and has the form
\begin{equation}\label{kinetic}
T_N= \sum_{i=1}^N T_i \ ,
\end{equation}
where $T_i$ acts on the coordinate of the $i^{th}$ electron.
The nonrelativistic choice is $T=p^2$ with $\p=-i\boldsymbol{\nabla}$
and $p^2 = -\uDelta$.

\subsection{Nonrelativistic Stability for Fermions}

The problem of stability of the second kind for nonrelativistic
quantum mechanics was recognized in the early days by a few
physicists, e.g., Onsager, but not by many. It was not solved until
1967 in one of the most beautiful papers in mathematical physics by
Dyson and Lenard \cite{dl}. 

They found that the Pauli principle, i.e., Fermi-Dirac statistics, is
essential.  Mathematically, this means that the Hilbert space is the
subspace of antisymmetric functions, i.e.,
${\mathcal{H}}^{\mathrm{phys}} =
\wedge^N  L^2({\mathbb{R}}^3; {\mathbb{C}}^2)$. This is how the 
Pauli principle is interpreted post-Schr\"odinger; Pauli invented his
principle a year earlier, however!

Their value for $C$ in (\ref{secondstab}) was rather high, about
$-10^{15}$ eV for $Z=1$. The situation was improved later by Thirring
and myself \cite{liebthirring} to about $-20$ eV for $Z=1$ by
introducing an inequality that holds only for the kinetic energy of
fermions (not bosons) in an arbitrary state $\Psi$.
\begin{equation}\label{lt}
\langle \Psi, T_N \Psi \rangle \geq (const.) \int_{\mathbb{R}^3}
\varrho_\Psi(\x)^{5/3} \, \D^3 \x \  ,
\end{equation}
where $\varrho_\Psi$ is the one-body density in the (normalized)
fermionic wave function $\Psi$ (of space and spin) given by an
integration over $(N-1)$ coordinates and $N$ spins as follows.
\begin{equation}\label{rho}
\varrho_\Psi(\x) = N\sum_{\sigma_1,\dots ,
\sigma_N} \int_{\mathbb{R}^{3(N-1)}}|\Psi(\x,\, \x_2,...,\x_N; 
\sigma_1,\dots \sigma_N)|^2\, \D^3\x_2\cdots \D^3\x_N  \  .
\end{equation} 

Inequality (\ref{lt}) allows one simply to reduce the quantum mechanical
stability problem to the stability of Thomas-Fermi theory, which was worked
out earlier by Simon and myself \cite{liebsimon}. 

The older inequality of Sobolev,
\begin{equation}\label{sobolev}
\langle \Psi, T_N \Psi \rangle \geq (const.)  \left(\int_{\mathbb{R}^3}
\varrho_\Psi(\x)^{3} \, \D^3 \x \right)^{1/3}\ ,
\end{equation}
is not as useful as (\ref{lt}) for the many-body
problem because its right side is proportional to $N$ instead of $N^{5/3}$. 

It is amazing that from the birth of quantum mechanics to 1967 none of
the luminaries of physics had quantified the fact that electrostatics
plus the uncertainty principle {\it do not suffice} for stability of
the second kind, and thereby make thermodynamics possible (although
they do suffice for the first kind). See Sect. \ref{bose}.  It was
noted, however, that the Pauli principle was responsible for the large
sizes of atoms and bulk matter (see, e.g., \cite{dyson1,dl}).

\subsection{Nonrelativistic Instability for Bosons}\label{bose}

What goes wrong if we have charged bosons instead of fermions?
Stability of the first kind (\ref{firststab}) holds in the
nonrelativistic case, but (\ref{secondstab}) fails. If we assume the nuclei 
are infinitely massive, as before, and $N=KZ$ then $E_0 \sim -N^{5/3}$
\cite{dl,lieb3}. To remedy the situation we can let the nuclei have finite
mass (e.g., the same mass as the negative particles). Then, as Dyson
showed \cite{dyson1}, $E_0 \leq -(const.)N^{7/5}$. This calculation
was highly non-trivial! Dyson had to construct a variational function
with pairing of the Bogolubov type in a rigorous fashion and this took
several pages.

Thus, finite nuclear mass improves the situation, but not enough. The question
whether $N^{7/5}$ is the correct power law remained open for many years. 
A lower bound of this type was needed and that was finally done in 
\cite{cly}.

The results of this Section \ref{nomagnet} can be summarized by saying
that stability of the hydrogen atom is one thing but stability of 
many-body physics is something else \thinspace !

\section{Relativistic Kinematics (no magnetic field)}\label{rel}

The next step is to try to get some idea of the effects of
relativistic kinematics, which means replacing $p^2$ by $\sqrt{p^2+
1}$ in non-quantum physics. The simplest way to do this is to substitute
$\sqrt{p^2+ 1}$ for $T$ in (\ref{kinetic}).  The Dirac operator will
be discussed later on, but for now this choice of $T$ will
suffice. Actually, it was Dirac's choice before he discovered his
operator and it works well in some cases. For example, Chandrasehkhar
used it successfully, and accurately, to calculate the collapse of
white dwarfs (and later, neutron stars).

Since we are interested only in stability, we may, and shall,
substitute $|\p| = \sqrt{-\uDelta}$ for $T$. The error thus introduced
is bounded by a constant times $N$ since $|\p|<\sqrt{p^2+ 1}< |\p|+1$
(as an operator inequality).  Our Hamiltonian is now $H_N
=\sum_{i=1}^N |\p_i| +\alpha V_c$.

\subsection{One-Electron Atom}\label{oneelectron}

The touchstone of quantum mechanics is the Hamiltonian for `hydrogen'
which is, in our case,
\begin{equation}\label{relhyd}
H= |\p| -Z\alpha/|\x| = \sqrt{-\uDelta} -Z\alpha/|\x| \ .
\end{equation}

It is well known (also to Dirac) that the analogous operator with
$|\p|$ replaced by the Dirac operator ceases to make sense when
$Z\alpha >1$.  Something similar happens for (\ref{relhyd}).
\iffalse
\begin{eqnarray}
E_0 =0 \qquad\qquad\qquad &if& \ Z\alpha \leq 2/\pi \nonumber \\
E_0 = -\infty \qquad\qquad\qquad &if& \ Z\alpha >2/\pi \ .
\end{eqnarray}
\fi
\begin{equation} \label{crit}
E_0= 
\begin{cases} 
0 &\text{if $Z\alpha \leq 2/\pi$;} \\
-\infty   &\text{if $Z\alpha >2/\pi$ .}  
\end{cases}
\end{equation}

The reason for this behavior is that both $|\p|$ and
$|\x|^{-1}$ scale in the same way. Either the first term in
(\ref{relhyd}) wins or the second does.

A result similar to (\ref{crit}) was obtained in \cite{eps}
for the free Dirac operator $D(0)$ in place of $|\p|$, but with 
the wave function
$\Psi$ restricted to lie in the positive spectral subspace of $D(0)$. 
Here, the critical value is $\alpha Z \leq (4\pi)/(4+ \pi^2) >2/\pi$.

The moral to be drawn from this is that relativistic kinematics plus
quantum mechanics is a `critical' theory (in the mathematical sense).
This fact will plague any relativistic theory of electrons and the
electromagnetic field -- primitive or sophisticated.

\subsection{Many Electrons and Nuclei}\label{relmanybody}

When there are many electrons is it true that the condition
$Z\alpha \leq const.$ is the only one that has to be considered?
The answer is no! One {\it also} needs the condition that 
$\alpha $ itself must be small, regardless of how small $Z$ might be.
This fact can be called a `discovery' but actually it is an overdue
realization of some basic physical ideas. It should have been
realized shortly after Dirac's theory in 1927, but it does  not
seem to have been noted until 1983 \cite{daubechieslieb}. 

The underlying physical heuristics is the following. With $\alpha $
fixed, suppose $ Z\alpha = 10^{-6}\ll 1$, so that an atom is stable,
but suppose that we have $2\times 10^{6}$ such nuclei. By bringing
them together at a common point we will have a nucleus 
with $ Z\alpha =2$ and one electron suffices to cause collapse into it.
Then (\ref{firststab}) fails. What prevents this from happening,
presumably, is the nucleus-nucleus repulsion energy which goes to 
$+\infty$ as the nuclei come together. 
But this repulsion energy is proportional to $(Z\alpha)^2/\alpha$ and, 
therefore, if we regard $Z\alpha$ as fixed we see that $1/\alpha $
must be large enough in order to prevent collapse.

Whether or not the reader believes this argument, the mathematical
fact is that there is a fixed, finite number $\alpha_c \leq 2.72 $ 
 (\cite{liebyau}) so that when $\alpha > \alpha_c$
(\ref{firststab}) fails for {\it every} positive $Z$ and for 
every $N\geq 1$ (with or without the Pauli principle). 

The open question was whether (\ref{secondstab}) holds for {\it all}
$N$ and $K$ if $Z\alpha $ and $\alpha $ are both small enough. The
breakthrough was due to Conlon \cite{conlon} who proved
(\ref{secondstab}), for fermions, if $Z=1$ and $\alpha <
10^{-200}$. The situation was improved by Fefferman and de la Lave
\cite{fl} to $Z=1$ and $\alpha < 0.16$.  Finally, the expected correct
condition $Z\alpha
\leq 2/\pi$ and $\alpha < 1/94$ was obtained in (\cite{liebyau}). (This
paper contains a detailed history up to 1988.) The situation was
further improved in (\cite{lls}). The multi-particle version of the use
of the free Dirac operator, as in Sect. \ref{oneelectron}, was treated 
in \cite{hs}.

Finally, it has to be noted that charged bosons are {\it always} unstable
of the first kind (not merely the second kind, as in the nonrelativistic
case) for {\it every} choice of $Z>0, \alpha > 0$. E.g., 
there is instability if  $Z^{2/3}\alpha N^{1/3} > 36$ (\cite {liebyau}). 

We are indeed fortunate that there
are no stable,  negatively charged  bosons.  

\section{Interaction of Matter with Classical Magnetic Fields}\label{magfields}

The magnetic field $\B$ is defined by a vector potential $\A(\x)$ and 
$\B(\x) =\mathrm{curl}\, \A(\x)$. In this section we take a first step 
(warmup exercise) by
regarding $\A$ as classical, but indeterminate, and we introduce the
classical field energy 
\begin{equation}\label{classfield}
H_f = \frac{1}{8\pi}\int_{\mathbb{R}^3} B(\x)^2 \D x \ .
\end{equation}

 The Hamiltonian is now
\begin{equation}\label{fieldham}
H_N(\A) = T_N(\A)+ \alpha V_c + H_f \ ,
\end{equation}
in which the kinetic energy operator has the 
form (\ref{kinetic}) but depends on $\A$.
We now define $E_0$ to be  the infimum of 
$\langle \Psi,\  H_N(\A) \Psi \rangle$ both with respect to
$\Psi $ {\it and with respect to} $\A$. 

\subsection{Nonrelativistic Matter with Magnetic Field}

The simplest situation is merely `minimal coupling' without spin, namely,
\begin{equation}
T(\A) = |\p +\sqrt\alpha \A(\x)|^2
\end{equation}
This choice does not change any of our previous results qualitatively.
The field energy is not needed for stability.  On the one particle
level, we have the `diamagnetic inequality' $\langle \phi,\ |\p+\A(\x)
|^2 \phi \rangle \geq \langle |\phi|,\ p ^2 |\phi| \rangle$. The same
holds for $|\p+\A(\x)|$ and $|\p|$. More importantly, inequality
(\ref{lt}) for fermions continues to hold (with the same constant)
with $T(\A)$ in place of $p^2$. (There is an inequality similar to
(\ref{lt}) for $|\p|$, with $5/3$ replaced by $4/3$, which also
continues to hold with minimal substitution \cite{daubechies}.)

The situation gets much more interesting if spin is included. This takes
us a bit closer to the relativistic case. The kinetic energy operator
is the Pauli-Fierz operator
\begin{equation}
T^P(\A) = |\p + \sqrt\alpha\; \A(\x)|^2 + \sqrt\alpha \;\B(\x)\cdot \s\ ,
\end{equation}
where $\s$ is the vector of Pauli spin matrices. 

\subsubsection{One-Electron Atom}

The stability problem with $T^P(\A)$ is complicated, even for a
one-electron atom.  Without the field energy $H_f$ the Hamiltonian is
unbounded below.  (For fixed $\A$ it is bounded but the energy tends
to $-\infty$ like $-(\log B)^2$ for a homogeneous field
\cite{ahs}.) The field energy saves the day, but the result is surprising 
\cite{fll}
(recall that we must minimize the energy with respect to $\Psi$ {\it and}
$\A$):
\begin{equation}
|\p + \sqrt\alpha\; \A(\x)|^2 + \sqrt\alpha \;\B(\x)\cdot \s
-Z\alpha/|\x| +H_f
\end{equation}
{\it is bounded below if and only if $Z\alpha^2 \leq C$,} where $C$ is some 
constant that can be bounded as $1<C <9\pi^2/8$. 

The proof of instability \cite{ly} is difficult and requires the
construction of a zero mode (soliton) for the Pauli operator, i.e., a
finite energy magnetic field and a {\it square integrable} $\psi$ such
that 
\begin{equation}
\label{zeromode} T^P(\A)\psi =0\ .  
\end{equation}
The usual kinetic energy
$|\p+\A(\x)|^2$ has no such zero mode for any $\A$, even when 0 is
the bottom of its spectrum.

The original magnetic field \cite{ly} that did the job in
(\ref{zeromode}) is independently interesting, geometrically (many
others have been found since then). 
$$ \B(x) = {12 \over (1 + x^2)^3}
[(1-x^2) \mathbf{w} + 2(\mathbf{w}\cdot \x) \x + 2 \mathbf{w}\land \x]
$$ 
with $\vert \mathbf{w} \vert = 1$. The field lines of this magnetic
field form a family of curves, which, when stereographically projected
onto the 3-dimensional unit sphere, become the great circles in what
topologists refer to as the Hopf fibration.

Thus, we begin to see that nonrelativistic matter with magnetic fields
behaves like relativistic matter without fields -- to some extent.

The moral of this story is that a magnetic field, which we might think
of as possibly self-generated, can cause an electron to fall into the
nucleus. The uncertainty principle cannot prevent this, not even for
an atom!

\subsubsection{Many Electrons and Many Nuclei}

In analogy with the relativistic (no magnetic field) case, we can see that
stability of the first kind fails if $Z\alpha^2$ {\it or} $\alpha$ are 
too large. The heuristic reasoning is the same and the proof is similar.

We can also hope that stability of the second kind holds
 if both $Z\alpha^2$ {\it and} $\alpha$ are small enough.
The problem is complicated by the fact that it is the field energy $H_f$ that
will prevent collapse, but there there is only one field energy while there
are $N\gg 1$ electrons.

The hope was finally realized, however. Fefferman \cite{feff} proved stability
of the second kind for  $H_N(\A)$ with the Pauli-Fierz $T^P(\A)$ for
$Z=1$ and ``$\alpha$ sufficiently small''. A few months later it
was proved \cite{llsolovej} for $Z\alpha^2 \leq 0.04$ and $\alpha \leq
0.06$.  With $\alpha =1/137$ this amounts to $Z\leq 1050$. This very
large $Z$ region of stability is comforting because it means that
perturbation theory (in $\A$) can be reliably used for this particular
problem.

Using the results in \cite{llsolovej}, Bugliaro, Fr\"ohlich and Graf
\cite{bfg} proved stability of the same nonrelativistic 
Hamiltonian -- but with an ultraviolet cutoff, quantized magnetic
field whose field energy is described below. (Note: No cutoffs are
needed for classical fields.)

There is also the very important work of Bach, Fr\"ohlich, and Sigal
\cite{bfs} who showed that this nonrelativistic Hamiltonian with
ultraviolet cutoff, quantized field {\it and} with sufficiently small
values of the parameters has other properties that one expects. E.g.,
the excited states of atoms dissolve into resonances and only the ground
state is stable. The infrared singularity notwithstanding, the ground
state actually exists (the bottom of the spectrum is an eigenvalue);
this was shown in \cite{bfs} for small parameters and in \cite{gll}
for all values of the parameters.

\section{Relativity Plus Magnetic Fields} \label{relmag}

As a next step in our efforts to understand QED and the many-body problem
we introduce relativity theory along with the classical magnetic field.

\subsection{Relativity Plus Classical Magnetic Fields} \label{relmagc}

Originally, Dirac and others thought of replacing $T^P(\A)$ by
$\sqrt{T^P(\A) +1} $ but this was not successful mathematically and does not
seem to conform to experiment. Consequently, we introduce the Dirac operator
for $T$ in (\ref{kinetic}), (\ref{fieldham})
\begin{equation}\label{dirac}
D(\A) = \da \cdot \p + \sqrt{\alpha}\ \da \cdot \A(\x)
 + \beta m \ ,
\end{equation}
where $\da $ and $\beta$ denote the $4\times 4$ Dirac matrices
and  $ \sqrt{\alpha}$ is the electron charge as before. (This notation
of $\da$ and $\alpha$ is not mine.) 
We take $m=1$ in our units. The Hilbert space for $N$ electrons is
\begin{equation}\label{oldh}
\mathcal{H} = \wedge^N L^2(\mathbb{R}^3; \mathbb{C}^4)\  .
\end{equation}

The well known problem with $D(\A) $ is that it is unbounded below, and so
we cannot hope to have stability of the first kind, even with $Z=0$.
Let us imitate QED (but without pair production or renormalization) by
restricting the electron wave function to lie in the positive 
spectral subspace of a Dirac operator.

Which Dirac operator?

There are two natural operators in the problem. One is $D(0)$, the
free Dirac operator. The other is $D(\A)$ that is used in the
Hamiltonian. In almost all formulations of QED the electron is defined
by the positive spectral subspace of $D(0)$. Thus, we can define
\begin{equation} \label{hphys}
\mathcal{H}^{\mathrm{phys}} = P^+\ \mathcal{H} = \uPi_{i=1}^N  
\pi_i  \, \mathcal{H} \  ,
\end{equation}
where $ P^+=\uPi_{i=1}^N \pi_i $, and $\pi_i$ is the projector
of onto the positive
spectral subspace of $D_i(0) = \da \cdot \p_i 
+ \beta m$, the free Dirac operator for the $i^{\mathrm{th}}$ electron.
We then restrict the allowed wave functions in the variational principle
to those $\Psi$ satisfying
\begin{equation}
\Psi = P^+\ \Psi  \qquad\quad i.e., \ \Psi \in  
\mathcal{H}^{\mathrm{phys}}   \  .
\end{equation}

Another way to say this is  that we replace the Hamiltonian
(\ref{fieldham}) by $P^+ \, H_N \, P^+$ on $\mathcal{H}$ and look 
for the bottom of its spectrum.

It turns out that this prescription leads to disaster! While the use
of $D(0)$ makes sense for an atom, it fails miserably for the
many-fermion  problem, as discovered in \cite{lss} and refined in
\cite{gt}. The result is: 

{\it For all $\alpha >0$ in (\ref{dirac}) (with or without the Coulomb
term $\alpha V_c$) one can find $N$ large enough so that $E_0=
-\infty$.}

In other words, the term $\sqrt{\alpha}\, \da\cdot \A$ in the Dirac operator
can cause an instability that the field energy cannot prevent.

It turns out, however, that the situation is saved if
one uses the positive spectral subspace of the Dirac operator $D(\A)$ 
to define an electron. (This makes the concept  of an electron $\A$ 
dependent, but when we make the vector potential into a dynamical quantity
in the next section, this will be less peculiar since there will be no
definite vector potential but only a fluctuating quantity.) 
The definition of the physical Hilbert space is as in (\ref{hphys}) but with
$\pi_i$ being the projector onto the positive subspace of the 
full Dirac operator
$D_i(\A) = \da \cdot \p_i + \sqrt{\alpha}\ \da \cdot \A(\x_i)
 + \beta m $. Note that these $\pi_i$ projectors commute with each other
and hence their product $P^+$ is a projector.

The result \cite{lss} for this model ((\ref{fieldham}) with the Dirac
operator and the restriction to the positive spectral subspace of $D(\A)$)
is reminiscent of the situations we have encountered
before:

{\it If $\alpha $ and $Z$ are small enough stability of the second kind
holds for this model.} 

Typical stability values that are rigorously established \cite{lss} are
$Z\leq 56$ with $\alpha =1/137$ or $\alpha \leq 1/8.2$ with $Z=1$.

\subsection{Relativity Plus Quantized Magnetic Field}

The obvious next step is to try to imitate the strategy of Sect.
\ref{relmagc} but with the quantized $\A$ field. This was done recently
in \cite{liebloss}.

\begin{equation}\label{apot}
\A(\x) = \frac{1}{2\pi} \sum_{\lambda=1}^2 \int_{|\bk|\leq \uLambda}
\frac{\vec{\varepsilon}_\lambda(\bk)}{\sqrt{|\bk|}} \Big[
a_\lambda(\bk) e^{i\bk\cdot \x} + a_\lambda^{\ast}(\bk) e^{-i\bk\cdot 
\x}\Big]
\D^3 \bk \  ,
\end{equation}
where $\uLambda$ is the ultraviolet cutoff on the wave-numbers $|\bk|$.
The operators $a_{\lambda}, a^{\ast}_{\lambda}$
satisfy the usual commutation relations
\begin{equation}
[a_{\lambda}(\bk), a^{\ast}_{\nu} (\mathbf{q})] = \delta ( \bk-\mathbf{q})
\delta_{\lambda, \nu}\ , ~~~ [a_{\lambda}(\bk), a_{\nu} (\mathbf{q})] =
0, \quad {\mathrm{etc}}
\end{equation}
and the vectors $\vec{\varepsilon}_{\lambda}(\bk)$ are two
orthonormal polarization vectors perpendicular to $\bk$ and to each other.

The field energy $H_f$ is now given by a normal ordered version of 
(\ref{classfield})
\begin{equation}\label{eq:fielden}
H_f = \sum_{\lambda=1,2} ~ \int_{\mathbb{R}^3} ~ |\bk|\ 
a_\lambda^{\ast}(\bk)a_\lambda(\bk) \D^3 \bk
\end{equation}

The Dirac operator is the same as before, (\ref{dirac}). Note that 
$D_i(\A)$ and $D_j(\A)$ still commute with each other (since
$\A(\x)$ commutes with $\A(\mathbf{y})$). This is important because it allows
us to imitate Sect. \ref{relmagc}.

In analogy with (\ref{oldh}) we define 
\begin{equation}
\mathcal{H} = \wedge^N L^2(\mathbb{R}^3; \mathbb{C}^4)
\otimes \mathcal{F}\ , 
\end{equation}
where $\mathcal{F}$ is the Fock space for the
photon field. We can then define the {\it physical} Hilbert space as before
\begin{equation}
\mathcal{H}^{\mathrm{phys}} = \Pi\ \mathcal{H} = \uPi_{i=1}^N  
\pi_i  \, \mathcal{H}\  ,
\end{equation}
where the projectors $\pi_i$ project onto the 
positive spectral subspace of either $D_i(0)$ or
$D_i(\A)$.

Perhaps not surprisingly, the former case leads to catastrophe, as before.
This is so, even with the ultraviolet cutoff, which we did not have
in Sect.  \ref{relmagc}. Because of the cutoff the catastrophe is milder
and involves instability of the second kind instead of the first kind.
This result relies on a coherent state construction in \cite{gt}.

The latter case (use of $D(\A)$ to define an electron)
leads to stability of the second kind if $Z$ and $\alpha $ are not 
too  large. Otherwise, there is instability of the second kind. 
The rigorous estimates are comparable  to the ones in
Sect. \ref{relmagc}.

Clearly, many things have yet to be done to understand the 
stability of matter in the context of QED. Renormalization and
pair production have to be included, for example.

The results of this section suggest, however, that a significant
change in the Hilbert space structure of QED might be necessary.
We see that it does not seem possible to keep to the current
view that the Hilbert space is a simple tensor product of a space for the 
electrons and a Fock space for the photons. That leads to
instability for many particles (or large charge, if the idea of 
`particle' is unacceptable). The `bare' electron is not really a good
physical concept and one must think of the electron as always accompanied
by its electromagnetic field. Matter and the photon field are inextricably
linked in the Hilbert space $\mathcal{H}^{\mathrm{phys}} $.

%\newpage
The following tables \cite{liebloss} summarize the results of this and
the previous sections
\bigskip
\bigskip

\centerline{\bf Electrons defined by projection onto the positive}
\centerline{\bf subspace of $D(0)$, the free Dirac operator}

\bigskip
%\begin{table}
%\quad\quad
\begin{tabular}{l||c|c|}%\cline{2-3}\cline{2-3}
&Classical or quantized field  & Classical or quantized field \\
 &\quad without cutoff $\uLambda$ &  with cutoff $\uLambda$ \\
& $\alpha >0$ but arbitrarily small. & $\alpha >0$ but arbitrarily small.\\
 & & \\
\hline\hline
Without Coulomb&  Instability of &  Instability  of \\
potential $\alpha V_c$  & the first kind& the second kind \\
 \hline
With Coulomb &  Instability of &  Instability of \\
potential $\alpha V_c$ & the first kind &  the second kind \\
 \hline\hline
\end{tabular}%\end{table}

\bigskip\bigskip
%\newpage
\vskip .4 true in

\centerline{\bf Electrons defined by projection onto the  positive}
\centerline{\bf subspace of $D(\A)$, the Dirac operator with field}

\bigskip
%\quad\quad 
\begin{tabular}{l||c|c|}%\hline\hline
&\multicolumn{2}{c| } {Classical field with or without cutoff $\uLambda$ } \\
&\multicolumn{2}{c| } {or quantized field with  cutoff $\uLambda$} \\
% &Classical or quantized field & Classical or quantized field \\
%q&\quad without cutoff $\uLambda$. & \quad with cutoff $\uLambda$. \\
   &\multicolumn{2}{c | }   {}  \\
\hline\hline
Without Coulomb& \multicolumn{2}{c |} {The Hamiltonian is positive}  \\
potential $\alpha V_c$ &  \multicolumn{2}{c |}  {}     \\
 \hline
&\multicolumn{2}{c |} {Instability of the first kind
when either} \\
With Coulomb & \multicolumn{2}{c|} {$\alpha$ or $Z\alpha$ is too large}\\
\cline{2-3}
potential $\alpha V_c$ & \multicolumn{2}{c|}{Stability of the second kind
when}\\
& \multicolumn{2}{c| }{both $\alpha$ and $Z\alpha$ are small enough}\\
\hline \hline
\end{tabular}

%\newpage

\end{document}